# Ripple-to-dome transition: the growth evolution of Ge on vicinal Si(1 1 10) surface.

L. Persichetti, A. Sgarlata, M. Fanfoni, and A. Balzarotti

*Dipartimento di Fisica, Università di Roma "Tor Vergata",*
*Via della Ricerca Scientifica,1-00133 Roma, Italy.*

We present a detailed scanning tunnelling microscopy study which describes the morphological transition from ripple to dome islands during the growth of Ge on the vicinal Si(1 1 10) surface . Our experimental results show that the shape evolution of Ge islands on this surface is markedly different from that on the flat Si(001) substrate and is accomplished by agglomeration and coalescence of several ripples. By combining first principle calculations with continuum elasticity theory, we provide an accurate explanation of our experimental observations.

The heteroepitaxial growth of Ge and SiGe islands on vicinal Si(001) substrates has attracted wide interest as a model system for exploiting self-organized nanoscale texturing on surfaces [1–4]. Although the main mechanisms involved in Ge growth on the flat Si(001) have been elucidated [5–7], the vicinal systems exhibit many remarkable features which are still only partially understood. For example, it is well-known that three-dimensional islands grown on the flat Si(001) surface surface show a bimodal behavior with shallow {105}-faceted pyramids (huts) at small volumes and steeper multifaceted domes at larger sizes [8, 9]. The pyramid-to-dome transition is driven by an abrupt change in chemical potential at a certain critical volume, corresponding to the crossover between the energy per atom of a dome and a corresponding energy per atom for a pyramid [9]. During this morphological transition, a single Ge pyramid progressively converts to dome by step-bunching at the island apex which generates new steeper facets [10]. By contrast, a recent experimental study of Ge growth as a function of substrate vicinality [11] revealed that on the Si(1 1 10) surface [(001) substrate misoriented ≈8° toward the [110] direction] pyramidal nano-sized islands transform into elongated nanoripples which are prisms of triangular cross section bounded by two adjacent {105} facets [12, 13]. Hitherto, the detailed pathway which leads to dome formation from ripple-like islands was unknown. Here, we report a systematic scanning tunnelling microscopy (STM) study which describes this morphological transition. Our results show that the shape evolution of Ge islands on this surface differs markedly from that on the flat (001) surface and is accomplished by agglomeration and coalescence of several ripples. We corroborate our analysis with a realistic calculation of the formation energy of multifaceted islands on the 8°-miscut Si(001) surface, in comparison with the flat case. By combining experimental observations and theoretical results, we extend the thermodynamic model for the formation of multifaceted islands on Si(001) to include the unconventional features of the growth on vicinal surfaces.

Experiments were carried out in an ultrahigh vacuum chamber (p<3x10⁻¹¹ torr) equipped with a variable temperature scanning tunneling microscope. The substrates were cleaned in situ by a standard flashing procedure at 1473 K [14]. Ge was deposited by physical vapor deposition at 873 K at constant flux of $(1.8\pm0.2)$x$10^{-3}$ ML/s (1ML corresponds to $6.3$x$10^{14}$ atoms per cm²). The flux was calibrated from the increasing area of terraces between two successive STM images during the layer-by-layer growth [15]. STM measurements were carried out at room temperature in the constant-current mode, using W-probe tips.

At a Ge coverage $\theta_1$=(4.0±0.2) ML, the Si(1 1 10) surface exhibits a composite morphology where different structures coexist [Fig 1(a)]. On a rough wetting layer (WL), {105}-faceted undulations are locally formed, resulting in an isolated unit (pre-ripple) [Fig. 1(b)]. Adjacent {105} facets grow bottom-up until they meet at the top of the pre-ripple and, subsequentely, extend laterally as the island elongates along the [110] direction [Fig. 1(c)]. Progressively, new {105} layers grow on the top of each other, producing the characteristic multilayered structure (ripple), depicted in Fig. 1(d). This structure is not symmetric, as evident from the enlarged views, shown in Fig. 1(e) and (f), which display the opposite ends of the island along the miscut direction. On one side [Fig. 1(e)], the ripple is not closed by any real facet, but gradually lowers in height and width as the number of the stacked {105}-layers decreases near the end of the island. On the opposite side [Fig. 1(f)], the closure is sharper and consists of growing facets oriented approximately perpendicular to the miscut direction. The overall morphology can be easily imaged as the result of cutting a {105}-pyramid with a (1 1 10) plane along the [110] direction, as schematically displayed in Fig. 2(i). Since the (1 1 10) plane is parallel to the [55$\bar{1}$] intersection line of two adjacent facets of the pyramid, the down side of the ripple cannot be bounded by {105} facets, confirming the experimental observations. On the Ge(105) surface, atoms form ordered arrays of U-shaped structures which are organized into zig-zag rows orthogonal to the [010] direction [rebonded-step (RS) reconstruction] [16–18]. This is the reason why, on the flat Si(001) surface, the {105}-side facets of the pyramids are oriented along the <010> directions. Thus, the rows of the RS-reconstruction are orthogonal to the pyramid edge (Fig. 2[a,c,e]). Since the vicinal (1 1 10) surface consists of arrays of (001) terraces separated by steps, in order to ensure a good matching to the WL, the rows must be kept orthogonal to the [010] direction. As a consequence, they form a 45° angle with the ripple edge, which is along the [110] direction [Fig. 2(b,d,f)]. It is worth noting that, due to their peculiar growth mode, most of the steps on the {105} ripple facets run parallel to the rows (i.e. along the [501] direction). A detailed analysis of these steps is still lacking, since most of the previous work was focused on the steps oriented along the [010] direction, which are relevant on the flat surface [17]. Therefore, we performed a systematic high-resolution STM study of such steps on the ripple facets. The typical configuration of a step is reported in Fig. 2(g). On the upper terrace [which is on the right hand side of Fig. 2(g)], the RS reconstruction shows structural changes around the step edge. In particular, the outermost atoms of the U-shaped structures pointing towards the step edge are rearranged. We also find that steps are mostly bunched together. The height of a step can be easily measured, since it is solely determined by the structural relationship between the upper and the lower terrace. Equivalent sites on the rows of the top and the bottom terraces are shifted along the [501] direction by a quantity $\Delta = n\ s_x$ for a step of $n$ monolayers ($s_x$=2.774 Å; 1 ML= 0.55 Å) [17]. From the analysis of the row profiles in adjacent terraces, the most common configuration on adjacent terraces is a doubled-layer step, in which the rows are almost in antiphase [Fig. 2(g,h)].

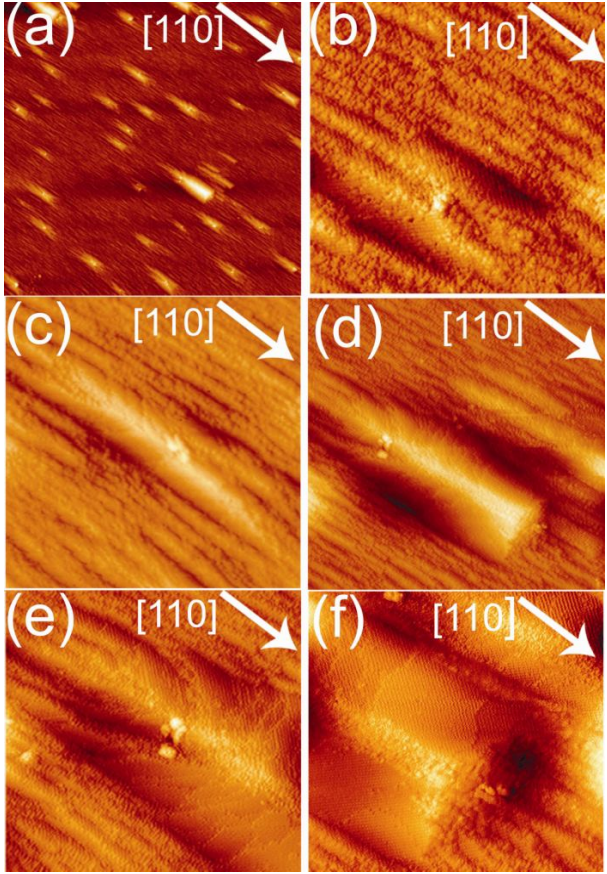

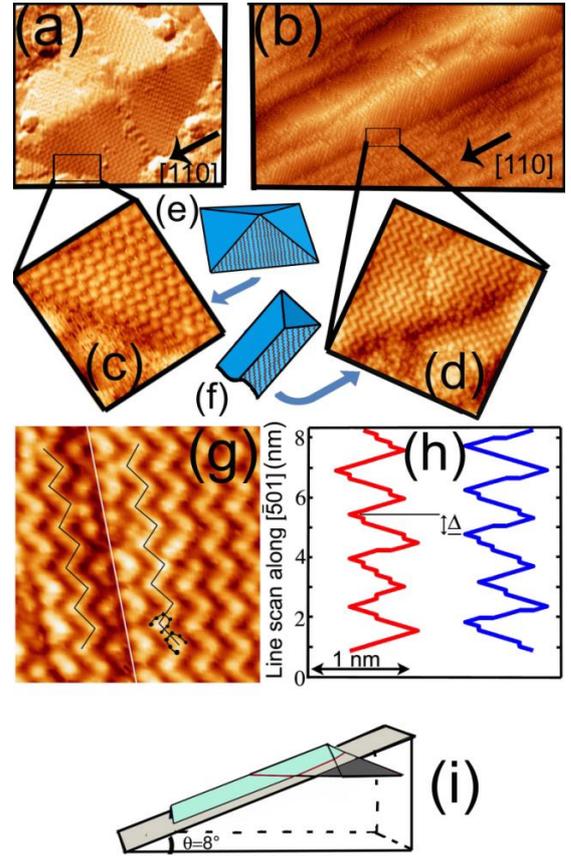

FIG. 1: (color online) STM images of the Si(1 1 10) surface with a (4.0 ± 0.2) ML of Ge coverage. (a) (950x950) nm², (b) (80x80) nm², (c) (150x150) nm², (d) (200x200) nm², (e) (120x120) nm².

FIG. 2: (color online) (a) (55x55) nm² STM image of a pyramidal hut on the flat Si(001) surface. (b) (88x75) nm² STM image of a ripple on the Si(1 1 10) surface. (c-d) Blow-up of Ge(105) RS-reconstruction near the island edges: (c) (14x13) nm² of a hut, (d) (20x20) nm² of a ripple. (e-f) Schematics showing the orientation of the rows of the RS-reconstruction for huts (e) and ripples (f). (g) Filled-state STM image (9.2x9.8) nm² of a [501] step on the {105} facet of a ripple ($V_S$=-1.85 V, I=0.85 nA). The step is highlighted by a straight line, and the rows by segmented lines. The U-shaped structures of the RS-reconstruction are superimposed onto the image. (h) Profiles measured in panel (g). (i) Geometry of a ripple as resulting from cutting a {105} pyramid with a (1 1 10) plane.

So far, we have described the detailed structure of Ge ripples on the vicinal Si(1 1 10) surface. We now go on to show that, at the same coverage $\theta_1$ at which these ripple-like structures form precursors of domes are present on the surface too. This is rather surprising, if compared with the growth on the flat surface, in which domes are formed at much higher coverage ($> 6$ ML) [10]. The sample surface displays several different stages on the shape transition to domes, as shown in Fig. 3(a-c). The morphological transformation starts with the local aggregation of ripples [Fig. 3(a)] and proceeds with the coalescence of the individual ripple units [Fig. 3(b)]. As the volume increases by aggregation and coalescence, the transition islands assume a rounded shape [Fig. 3(c)]. At this point, a further evolution is attained either by an annealing step at 993 K for 10 min [Fig. 3(d)] or by a slight increase of the Ge coverage up to $\theta_2=(4.8\pm0.2)$ ML [Fig. 3(e)]. The resulting morphologies are similar to each other and to the final dome shape. However, upon deposition of additional Ge, the evolution rate and the number of transitional islands is increased.

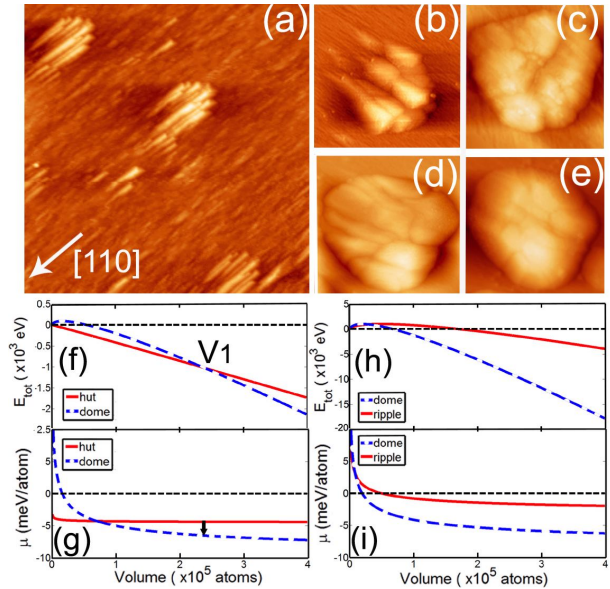

FIG. 3: (color online) (a-c) STM images illustrating different steps on the transition to domes at $(4.0 \pm 0.2)$ ML of coverage. Morphological modifications due to: (d) 10 min annealing at 993 K, (e) additional deposition of $\sim 1$ ML of Ge. (f-i) Thermodynamic stability of Ge islands on the flat and vicinal substrate. (f) Total energy vs volume for huts and domes on the flat surface and (f) the corresponding behavior of chemical potentials. (h) Total energy and (i) chemical potential of ripples and domes on the Si(1 1 10) surface.

In the following, we model the essential features of the transition to domes on the vicinal surface with respect to the substrate. It is well-known that, disregarding the edge contribution for a large island volume, the total energy gain associated to the formation of a 3D island of volume $V$ on the WL is

$$E_{tot} = e_{relax}V + e_{surf}V^{2/3}. \qquad (1)$$

The first term represents the bulk strain relief and the second one accounts for the formation of island facets. The energy density of elastic relaxation, $e_{relax}$, is the (negative) difference between the residual strain energy stored in a Ge island of volume $V$ and in the Si substrate after relaxation and the energy in an equivalent volume $V$ of a fully strained epitaxial Ge film. The surface term, $e_{surf}$, is the extra surface energy per unit area due to the presence of the island

$$e_{surf} = \left[ \sum_{i=1}^{N} \gamma_i \left( \varepsilon_i \right) S_i - \gamma_0 S_0 \right] V^{-2/3}, \qquad (2)$$

where $\gamma_i(\varepsilon_i)$ and $S_i$ denote the surface energy and surface area of the i-th facet of the island. $S_0$ is the base area of the island, and $\gamma_0$ is the energy per unit area of the WL. In order to estimate $E_{tot}$, we treat the elastic term within the continuum elasticity theory, using finite element method (FEM) calculations to evaluate the elastic energy relaxation. For the surface term, we use published density functional theory (DFT) data for the surface energies, whenever available. For the flat Si(001) case, we set $\gamma_0$ to a value of 60.4 meV/Å$^2$, corresponding to the energies of Ge/Si(001) with 4 layer of Ge [19, 20], after subtracting the Ge/Si interfacial energy ($\approx$1 meV/Å$^2$, see Ref. [21]). The surface energy of the vicinal Si(1 1 10) surface, is estimated as $\gamma_0(\theta) = \gamma_0 \cos \theta + \beta \sin \theta$, where $\theta=8°$ is the miscut angle and $\beta$ is the step formation energy per unit height [22]. Since the {105}-facets appear at the base of both pyramids and ripples and hence may be highly strained, we take into account the strain-dependent correction to the surface energies of these facets. This is done by interpolating *ab initio* results for the dependence of the surface energy on strain [16] with the in-plane components of the strain on each facet, taken from FEM calculations. The resulting surface energies are 59.8 and 61.0 meV/Å$^2$ for the {105}-facets on pyramids and ripples, respectively. The surface energy of some of the facets of the domes has not yet been assessed by DFT. However, a previous analysis shows that an average value of 65 meV/Å$^2$ is a reasonable guess [23]. We verified that variations in the 61 to 69 meV/Å$^2$ range do not alter our findings appreciably.

Figure 3(f) shows the dependence of $E_{tot}$ vs $V$ for huts and domes on the flat Si(001) surface. The island's chemical potential, $\mu$, [Fig. 3(g)] is obtained by differenting Eq.1 with respect to the number of atoms (proportional to $V$) [24]. The curves follow the usual morphological evolution of the system at the growth temperature [9, 23]. Pyramids are always more stable than WL, resulting in a barrierless island formation. The shape transition to domes is energetically favored for volumes larger than the critical value $V_1$, where the energy curves cross. Moreover, the transition is first-order, since there is a discontinuity in $\mu$ (marked by an arrow) at the volume $V_1$ at which the two shapes are degenerate. In Fig. 3(h) and 3(i), we report analogous calculations for ripples and domes on the vicinal surface. While the dome behavior is almost unchanged, ripples become stable with respect to the WL at a volume at which domes already have a much lower energy $E_{tot}$. The smaller thermodynamic stability of ripples is due to their high surface-

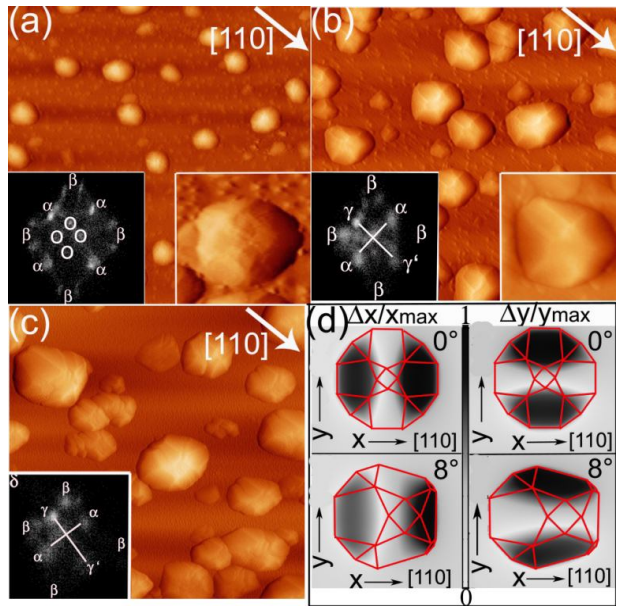

to-volume ratio, and to their lower elastic relaxation in comparison with huts [11]. Besides, the strain relaxation is concentrated along the lateral facets [25] and, thus, increases the local surface energy. According to the present analysis, domes are expected to nucleate when $E_{tot}$ becomes negative and, thus, at smaller volumes than on the flat substrate, perfectly matching the experiment. Even though domes have the lowest energy, kinetic considerations indicate that a large multifaceted structure is not likely to form directly from a film. Our experimental observations show that ripples act as precursors by collecting and piling up enough material into a dome shape. This resembles what has been found on flat Si(001) at

T>675°C [23] , where stable domes arise from the fluctuations of metastable pyramids.

Finally, we analyze the evolution of domes at large Ge coverage. Figure 4(b) shows the morphology of domes on the vicinal Si(1 1 10) surface, at a coverage of (7.0±0.2) ML. The corresponding surface orientation map [26] reveals that domes are "topologically asymmetric". In comparison with the domes on the flat surface, which have two symmetric {113} facets along the [110] direction, the domes on the vicinal surface display different facets on the opposite sides. The same morphology is observed on the Si(118) surface [(001) substrate misoriented ≈10° in the [110] direction] [Fig. 4(c)], indicating that the asymmetry is an intrinsic feature of domes nucleated on highly misoriented substrates. Recently, Spencer and Tersoff [22] have theoretically predicted that asymmetric Ge island shapes would occur on Si(001) at large miscut angles. Our experimental data confirm their findings. Moreover, our FEM calculations, made on 3D islands, show that the anisotropic shapes reflect the anisotropy of the elastic displacement field along the miscut direction. In Fig. 4(d), we report the relative displacement of domes along the miscut direction, $x$, and in the orthogonal direction. It can be seen that the displacement field is isotropic on the flat surface and anisotropic on the 8°-miscut substrate.

In summary, we have shown that the morphological evolution of Ge on vicinal Si(1 1 10) differs markedly from usual path on the flat (001) surface. By combining first-principle calculations and continuum modelling, we provide a consistent explanation of our experimental results.

This work has been supported by the MIUR-PRIN 2007 project No. 200754FAA4 of the Italian Ministry of Research.

---